\documentclass[a4paper]{article}
\usepackage[british]{babel}
\usepackage[utf8]{inputenc}
\usepackage{microtype}
\usepackage{amssymb}
\usepackage{amsthm}
\usepackage{amsmath}
\usepackage{graphicx}
\usepackage{color}
\usepackage{tikz}
\usepackage{here}
\usepackage{enumitem}
\usepackage{hyperref}
\usepackage{fp-eval}
\usepackage{subfigure}
\usepackage{booktabs}
\usepackage{caption}
\usepackage{framed}
\usepackage{ulem}
\usepackage{stackrel}
\usepackage{algorithm2e}
\usepackage[amssymb,mediumqspace]{SIunits}
\usepackage{authblk}

\graphicspath{{./} {../pics/}}

  \title{Induced charge electroosmotic flow with finite ion size and solvation effects}

\author[1]{J.~Fuhrmann}
\author[2]{C.~Guhlke}
\author[3]{C.~Merdon}
\author[4]{A.~Linke}
\author[5]{R. M\"uller}

\affil[1,2,3,4,5]{Weierstrass Institute, Mohrenstr. 39, 10117 Berlin, Germany}
\affil[1]{Juergen.Fuhrmann@wias-berlin.de}
\affil[2]{Clemens.Guhlke@wias-berlin.de}
\affil[3]{Christian.Merdon@wias-berlin.de}
\affil[4]{Alexander.Linke@wias-berlin.de}
\affil[5]{Ruediger.Mueller@wias-berlin.de}

\begin{document}

\maketitle

\begin{abstract}
The most common mathematical models for electrolyte flows are based on the dilute solution assumption, leading to a coupled system of the
Nernst--Planck--Poisson drift-diffusion equations for ion transport and the Stokes resp. Navier--Stokes equations for fluid flow of the solvent.
In charged boundary layers the dilute solution assumption is in general not valid
and volume exclusion and solvation effects have to be taken into account in a
thermodynamically consistent way.
Whenever boundary layer effects have a dominant impact on the global behavior of a
certain electrochemical system,
an accurate numerical simulation depends on the correct incorporation of these effects.
In this contribution we present a novel numerical solution approach which aims at preserving on the discrete level consistency with basic thermodynamic principles and structural properties like independence of flow velocities from gradient contributions to external forces.
We illustrate capabilities of the method by an example of vortex generation due to induced charge electroosmotic forces at an electrode inside a nanochannel.
\end{abstract}

\section{Introduction}
With the miniaturization of electrochemical devices and the trend towards nanometer scale electrochemistry \cite{ying2017advanced,stevenson2017materials,lin2018understanding}, understanding  the behavior of liquid electrolytes at the micro- and nanoscale becomes increasingly important.
Of particular interest are correct double layer models, the behavior of electroosmotic flows  at the nanoscale and their interaction with species transport.
Due to the complex physical interactions present in electrolyte flows, numerical simulation techniques are important tools for developing a deeper understanding of the flow behavior.

This contribution introduces a simulation approach for coupled ion transport and electrolyte flow.
Section \ref{sec:continuum} introduces a modified Nernst--Planck--Poisson--Navier--Stokes model which has its foundations in first principles of nonequilibrium thermodynamics  \cite{dreyer2013overcoming, dreyer2014mixture,landstorfer2016theory} and takes into account ion-solvent interactions, finite ion size and solvation effects.

Section \ref{sec:numerical}  introduces a finite volume discretization approach for ion transport in a self-consistent electric field which is motivated by results from semiconductor device simulation \cite{fuhrmann2016numerical}.
Pressure robust mixed finite element methods for fluid flow \cite{linke2014role}, and a fix point approach for coupling to ion transport are introduced.
%


Section \ref{sect:float} provides a numerical example which shows the potential significance of the proposed approach for modeling electrochemical processes at the nanoscale.

\section{The model}\label{sec:continuum}

Electroosmotic flows are characterized by the presence of an electric field that exerts a net force on the fluid molecules in regions where the local net charge due to the present ions  is nonzero.
Being part of the momentum balance for the barycentric velocity of the fluid, this net force induces fluid motion.
Correspondingly, ions are advected by the barycentric velocity field.
Their motion relative to the barycentric velocity is induced by the gradients of their chemical potential and the electrostatic potential.
The elastic interactions between the ions and the solvent exerts a counterforce against ion motion.

The ion charge density varies with the redistribution of the ions and contributes to the electric field.

The \textit{dilute solution} assumption lying at the foundation of classical electrolyte models \cite{nernst1888kinetik,planck1890ueber}  assumes that the ion concentration in the electrolyte is small  compared to the concentration of the solvent and consequently neglects the ion-solvent interaction.
However, accumulation of ions in  polarization boundary layers violates this assumption.
As a consequence, the resulting electrolyte model e.g. severely overestimates double layer  capacitances at ideally polarizable electrodes \cite{BardFaulkner}.

The model used throughout this paper  has been introduced in \cite{dreyer2013overcoming, dreyer2014mixture, landstorfer2016theory} based on consistent application of the principles of nonequilibrium thermodynamics \cite{deGrootMazur1962}.
It includes ion volume and solvation effects and consistently couples the transport equations to the momentum balance.
It generalizes previous approaches to include steric (ion size) effects \cite{bikerman1942,freise1952theorie,carnahan1969equation,mansoori1971equilibrium,kornyshev1981conductivity}, see also \cite{kilic2007steric}.

\subsection{Model equations}
In a given bounded space-time domain $\Omega\times (0,t^\sharp)\subset \mathbb R^d\times (0,\infty)$, and with appropriate initial and boundary conditions,
system \eqref{eq:balances}--\eqref{eq:constitutivefunc} describes the isothermal evolution of the molar concentration of $N$ charged species with molar densities (concentrations) $c_1\dots c_N$ with charge numbers $z_1\dots z_N$ dissolved in a solvent of concentration $c_0$.
Species are further characterized by their molar volumes $v_i$ and molar masses $M_i$.
The electric field is described as the gradient of the electrostatic potential $\psi$.
The barycentric velocity of the mixture is denoted by $\vec u$, and $p$ is the pressure.
The following equations are considered:
\begin{subequations}\label{eq:balances}
  \begin{align}
\rho \partial_t \vec u    -\nu\Delta \vec{u} + {\rho}(\vec{u} \cdot \nabla) \vec{u} + \nabla p    &= {q\nabla \psi}  \label{eq:moment}\\
    \nabla \cdot (\rho \vec{u})   &=0 \label{eq:div0} \\
    \partial_t c_i  + \nabla \cdot (\vec N_i + c_i {\vec{u}})  &=0 & (i=1\dots N) \label{eq:cont}\\
    -\nabla \cdot (\varepsilon \nabla \psi)    &= q. \label{eq:poisson}
  \end{align}
\end{subequations}

Equation \eqref{eq:moment} together with \eqref{eq:div0} comprises the incompressible Navier--Stokes equations for a fluid of viscosity $\nu$ and constant density $\rho$.
In the general case, where molar volumes and molar masses are not equal, $\rho$ would depend on the local composition of the electrolyte.

In regions where the charge density $q=F\sum_{i=1}^N z_i c_i$ ($F$ being the Faraday constant) is nonzero, the electric field $-\nabla \psi$
becomes a driving force of the flow.

The partial mass balance equations \eqref{eq:cont} describe the redistribution of species concentrations due to advection in the velocity field
$\vec u$ and molar diffusion fluxes $\vec N_i$.
The Poisson equation \eqref{eq:poisson} describes the distribution of the electrostatic potential $\psi$ under a given configuration of the charge density $q$.
The constant $\varepsilon$ is the dielectric permittivity of the medium.

The fluxes $\vec N_i$, the molar chemical potentials $\mu_i$ and the incompressibility constraint for a liquid electrolyte are given by
\begin{subequations}\label{sys:dgml}
  \begin{align}
    \vec N_i
    &= - \frac{D_i}{RT} c_i \left( \nabla \mu_i - \frac{\kappa_i M_0+M_i}{M_0}\nabla \mu _0 + z_i F \nabla \psi \right)
      & (i=1\dots N) \label{eq:npfull}\\
    \mu_i
    &=  (\kappa_i v_0+v_i)(p-p^\circ) + RT \ln \frac{c_i}{\overline c}
      & (i=0\dots N) \label{eq:constfull}\\
    1 &= v_0 c_0 + \sum_{i=1}^{N} ( \kappa_i v_0  + v_i)c_i.\label{eq:incompressfull}
  \end{align}
\end{subequations}
The modified Nernst--Planck flux \eqref{eq:npfull} combines the gradients of the species chemical potentials $\nabla \mu_i$, the gradient of the  solvent chemical potential $\nabla \mu_0$ and the electric field $-\nabla \psi$ as driving forces for the motion of ions of species $i$ relative to the barycentric velocity $\vec u$.
In this equation, $D_i$ are the diffusion coefficients, $R$ is the molar gas constant, and $T$ is the temperature.
Equation \eqref{eq:constfull} is a constitutive relation for the chemical potential $\mu_i$ depending on the local pressure and concentration.
Here, $p^\circ$ is a reference pressure, and $\overline c=\sum_{i=0}^N c_i$ is the  total species concentration.
In \eqref{eq:incompressfull} a simple model for solvated ions is applied, see  \cite{dreyer2014mixture,fuhrmann2016numerical,dreyer2017new}, which describes  the mass and volume of a solvated ion  by $\kappa_i M_0+M_i$ and $\kappa_i v_0+v_i$, respectively.
The incompressibility constraint \eqref{eq:incompressfull} limits the accumulation of ions in the polarization boundary layer to physically reasonable values \cite{dreyer2014mixture,fuhrmann2016numerical}.
%

The mass density of the mixture is
\begin{align}
      \rho&=M_0 c_0 + \sum_{i=1}^{N}(\kappa_i M_0+M_i) c_i. 
      \label{eq:rho}
\end{align}
As for reasonable solvation numbers $\kappa_i\approx 10$, molar masses and molar volumes of the solvated ions are dominated by the solvent mass and volume.
Therefore we assume, for simplicity, $(\kappa_i M_0+M_i)\approx (\kappa_i+1)M_0$ and $(\kappa_i v_0 + v_i) \approx (\kappa_i+1)v_0$.
This approximation leads to a simplified constitutive function and, in particular, to a constant mass density of the mixture,
\begin{subequations}\label{eq:constitutivefunc}
  \begin{align}
    \vec N_i
    &= - \frac{D_i}{RT} c_i \left( \nabla \mu_i - (\kappa_i+1)\nabla \mu _0 + z_i F \nabla \psi \right)
      & (i=1\dots N) \label{eq:np}\\
    \mu_i
    &=  v_0(\kappa_i+1)(p-p^\circ) + RT \ln \frac{c_i}{\overline c}
      & (i=0\dots N) \label{eq:const}\\
    1 &= v_0 c_0 + \sum_{i=1}^{N} ( \kappa_i  + 1)v_0c_i\label{eq:incompress}\\
    \rho &= \frac{M_0}{v_0}.\label{eq:constmass}
  \end{align}
\end{subequations}

Comparing the constitutive equations \eqref{eq:np}-\eqref{eq:incompress} to the classical Nernst--Planck flux \cite{nernst1888kinetik,planck1890ueber}
\begin{align}\label{eq:classflux}
  \vec N_i &= - D_i \left(\nabla c_i + z_i c_i \frac{F}{RT} \nabla \psi \right) & (i=1\dots N),
\end{align}
which considers dilute solutions, one observes  that in \eqref{eq:classflux} the ion-solvent interaction described by the term $\nabla \mu_0$ is ignored.
Moreover in \eqref{eq:classflux} implicitly a material model is assumed that neglects the pressure dependence of $\mu_i$, which is inappropriate in charged boundary layers \cite{dreyer2013overcoming}.


\subsection{Reformulation in species activities}

In order to develop a space discretization approach for the Nernst--Planck fluxes \eqref{eq:np}, after \cite{Fuhrmann2015xCPC}, the system is reformulated in terms of (effective) species activities $a_i= \exp\left(\frac{ \mu_i-(\kappa_i+1)\mu_0}{RT}\right)$.
The quantity $\mu_i - (\kappa_i+1)\mu_0$ is sometimes denoted as entropy variable \cite{jungel2015boundedness}.
Introducing the activity coefficients $\gamma_i=\frac{a_i}{c_i}$ and its inverse (reciprocal) $\beta_i=\frac1{\gamma_i}=\frac{c_i}{a_i}$ allows to transform the Nernst--Planck--Poisson system consisting of \eqref{eq:cont}, \eqref{eq:poisson}, \eqref{eq:np} to
\begin{subequations}\label{sys:act}
\begin{align}
    -\nabla \cdot (\varepsilon \nabla \psi)    &= F\sum_{i=1}^N z_i \beta_i a_i= q. \label{eq:apoisson}\\
    0 &=\partial_t (\beta_i a_i)  + \nabla \cdot (\vec N_i + \beta_i a_i {\vec{u}})   & (i=1\dots N) \label{eq:acont}\\
    \vec N_i&= -D_i \beta_i\left(\nabla a_i + a_i z_i \frac{F}{RT}\nabla\psi\right)
                                                                                & i=(1\dots N). \label{eq:actflux}
\end{align}
\end{subequations}
From \eqref{eq:const} and \eqref{eq:incompress} one obtains
\begin{align*}
  a_i&=\frac{v_0 \beta_i a_i}{1- v_0\sum_{j=1}^N \beta_j a_j (\kappa_j+1)} & (i=1\dots N)
\end{align*}
which is  a linear system of equations which allows to express $\beta_1\dots \beta_n$ through $a_1\dots a_n$.
It has the unique solution \cite{Fuhrmann2015xCPC}
\begin{align}\label{eq:beta}
  \beta_i=\beta=&\frac{1}{v_0+v_0\sum_{j=1}^Na_j(\kappa_j+1)}& (i=1\dots N).
\end{align}
It follows immediately that for any nonnegative solution $a_1\dots a_n$ of system \eqref{sys:act}, the resulting concentrations are bounded in a physically meaningful way:
\begin{align}\label{eq:cbound}
0\leq c_i=\beta_i a_i\leq \frac1{v_0}.
\end{align}

In the general case with different molar volumes and molar masses, system \eqref{eq:beta} becomes nonlinear, the quantities $\beta_i$ differ between species and in addition depend on the pressure $p$ \cite{Fuhrmann2015xCPC, fuhrmann2016numerical}, leading to a nonlinear system of equations
\begin{align}\label{eq:nlactcoeff}
      \beta_i&=B_i(a_1\dots a_n, \beta_1\dots \beta_n,p) & (i=1\dots N).
\end{align}

\section{A coupled Finite-Volume-Finite-Element discretization}\label{sec:numerical}

\subsection{Thermodynamically consistent finite volume methods for species transport }\label{subsec:fvol}

A two point flux finite volume method on boundary conforming Delaunay meshes is used to approximate the Nernst--Planck--Poisson part of the problem.
It has been inspired by the successful Scharfetter--Gummel box method for the solution of charge transport problems in semiconductors \cite{scharfetter1969large,bank1983numerical}.
For a recent overview on this method see \cite{farrell2017numerical}.
It was initially developed for drift-diffusion problems in non-degenerate semiconductors exhibiting Boltzmann statistics for charge carrier densities whose fluxes are equivalent to the classical Nernst--Planck flux \eqref{eq:classflux}.

The time axis is subdivided into intervals
$
  0=t^0< t^1 <\dots < t^{N_t}=t^\sharp.
$
The simulation domain $\Omega$ is partitioned into a finite number of closed convex polyhedral control volumes $K\in\mathcal K$ such that $K\cap L= \partial K\cap \partial L$ and $\overline\Omega=\bigcup_{K\in \mathcal K} K$.
With each control volume a node $\vec{x}_K\in K$ is associated. If the control volume intersects with the boundary $\partial \Omega$, its corresponding node shall be situated on the boundary: $\vec{x}_K\in\partial\Omega\cap K$.
The partition shall be \textit{admissible} \cite{EGH}, that is for two neighboring control volumes $K, L$, the \textit{edge} $\overline{\vec{x}_K\vec{x}_L}$ is orthogonal to the interface between the control volumes $\partial K \cap \partial L$.
Let $\vec h_{KL} = \vec x_L -\vec x_K$ and $h_{KL}=|\vec h_{KL}|$.
Then, the normal vectors to $\partial K$ can be calculated as $\vec n_{KL}=\frac1{h_{KL}}\vec h_{KL}$.

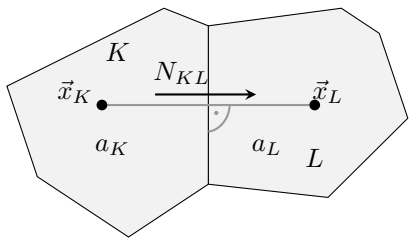
\begin{figure}
\centering
\begin{tikzpicture}[scale=0.7,>=stealth,shorten >=2pt]
 \filldraw[fill=gray!10, draw=black]
( 2.00000, 1.50000) --
( 4.50000, 1.83333) --
( 5.21429, 1.35714) --
( 5.75000,-0.25000) --
( 4.25000,-1.75000) --
( 2.00000,-1.50000) --
 cycle;
 \filldraw[fill=gray!10, draw=black]
( 2.00000, 1.50000) --
( 1.16667, 1.83333) --
(-1.78571, 0.35714) --
(-1.21429,-1.35714) --
( 0.50000,-2.50000) --
( 2.00000,-1.50000) --
 cycle;
 \draw[thick,gray!80] (0,0) -- (4,0);
\draw (0,0) node[inner sep=0.5mm,shape=circle,fill] {} ;
\draw (4,0) node[inner sep=0.5mm,shape=circle,fill] {} ;
\draw (-0.5,0.25) node {$\vec{x}_K$};
\draw (4.25,0.25) node {$\vec{x}_L$};
\draw(0.3,1) node {$K$};
\draw(0.2,-0.8) node {$\textcolor{black}{a_{K}}$};
\draw (4,-1) node {$L$};
\draw(3.1,-0.8) node {$\textcolor{black}{a_{L}}$};
\draw[thick,black,->] (1,0.2) -- (3,0.2);
\draw(1.5,0.6) node {$\textcolor{black}{N_{KL}}$};
\draw [thick, gray!80] (2.4,0.0) arc (360:270:0.5);
\draw (2.15,-0.15) node[thick,gray!80,inner sep=0.2mm,shape=circle,fill] {};
\end{tikzpicture}
  \caption{Two neighboring control volumes $K$ and $L$ with
    collocation points $\vec x_K, \vec x_L$
    stored activities $a_K, a_L$ and flux $N_{KL}$.\label{FIG:sgfv}}
\end{figure}

A constructive way to obtain such a partition is based on the creation of a boundary conforming Delaunay triangulation resp. tetrahedralization of the domain and the subsequent construction of its dual, the Voronoi tessellation intersected with the domain, see e.g. \cite{bank1983numerical,SiFuhrmannGaertner2010,farrell2017numerical},
see also Figure \ref{FIG:sgfv}.

Denote by $\vec J_i= c_i\vec u + \vec N_i = \beta_i a_i\vec u + \vec N_i$ the convection diffusion flux of the model under consideration.
The general approach to derive a two point flux finite volume scheme for a conservation law
(index $i$ omitted)
\begin{align*}
  \partial_t c + \nabla\cdot \vec J=0
\end{align*}
consists in integrating the equation over a space-time control volume $K\times [t^{n-1},t^n]$:
\begin{align*}
  0
  = & \int\limits_{K} (c^n-c^{n-1}) \ d\vec x  + \sum\limits_{\substack{L\; \text{neighbor}\\\text{ of}\; K}}\ \int\limits_{t^{n-1}}^{t^n}\int\limits_{\partial K\cap \partial L}\vec J\cdot \vec n_{KL}\ ds \ dt.
\end{align*}
This is approximated via
\begin{align*}
  0 = \lvert K \rvert\left(c_K^n-c_K^{n-1} \right)+
  \sum\limits_{\substack{L\; \text{neighbor of}\; K}}
  \left(t^n-t^{n-1}\right) \lvert\partial K\cap\partial L\rvert \, J^n_{KL},
\end{align*}
and it remains to define the numerical fluxes $J^n_{KL}$ which should approximate the continuous fluxes between two neighboring control volumes and depend on the unknown values in the two collocation points $\vec x_K$ and $\vec x_L$ at moment $t^n$ in order to obtain a fully implicit in time scheme.

The modification of the Scharfetter-Gummel scheme \cite{scharfetter1969large} proposed in \cite{Fuhrmann2015xCPC} is based on the similarity of the expressions \eqref{eq:actflux} and \eqref{eq:classflux}.
The flux expression  \eqref{eq:classflux} is the same as the drift-diffusion flux in non-degenerate semiconductors, for which this discretization scheme was initially derived.
The only difference between \eqref{eq:actflux} and \eqref{eq:classflux}  is the multiplication with the pre-factor $\beta$.
Therefore it appears to be reasonable to mimic this structure at the discrete level, and to derive a discrete equivalent of \eqref{eq:actflux} from the discrete version of \eqref{eq:classflux} by multiplication with a suitable average of $\beta$: set
\begin{align*}
   J_{KL}=D\,\beta_{KL}\cdot\frac{B\left(-\delta_{KL} - \frac{u_{KL}}{D} \right)a_{K}-B\left(\delta_{KL}+\frac{u_{KL}}{D}\right)a_{L}}{h_{KL}},
\end{align*}
where $B(\xi)=\frac{\xi}{e^\xi-1}$ be the Bernoulli function.
and $\beta_{KL}$ is an average of the inverse activity coefficients $\beta_K$ and $\beta_L$, $\delta_{KL}=\frac{zF}{RT}(\psi_K -\psi_L)$ is proportional to the local electric force, and
\begin{align}\label{eq:fluxint}
  u_{KL} & = \int_{\partial K \cap \partial L} \vec u \cdot \vec h_{KL} \, ds
\end{align}
is the normal integral over the interface  $\partial K \cap \partial L$ of the convective flux scaled by $h_{KL}$.
If the continuous flux is divergence free, i.e. it fulfills equation \eqref{eq:div0}, the flux projections $u_{KL}$ fulfill the discrete divergence condition
\begin{align}\label{eq:discdiv0}
  \sum\limits_{\substack{L\; \text{neighbor of}\; K}} \lvert \partial K \cap \partial L \rvert \, u_{KL}=0
\end{align}
which in the absence of charges and coupling through the activity coefficients guarantees a discrete
maximum principle of  the approximate solution \cite{fuhrmann2011numerical}.

The resulting time discrete implicit Euler finite volume upwind scheme guarantees nonnegativity of discrete activities and exact zero fluxes under thermodynamic equilibrium conditions.
Moreover it guarantees the bounds \eqref{eq:cbound} \cite{Fuhrmann2015xCPC}.
%

%

\subsection{Pressure  robust, divergence free finite  elements for  fluid flow.}\label{subsec:pr}

Mixed finite element methods approximate the Stokes resp. Navier--Stokes equations based on  an inf-sup stable pair of velocity ansatz space $V_h$ and pressure ansatz space $Q_h$.
A fundamental property of the Stokes and Navier--Stokes equations consists in the fact that --- under appropriate boundary conditions --- the addition of a gradient force to the body force on the right-hand side of the momentum balance \eqref{eq:moment} leaves the velocity unchanged, as it just can be compensated by a change in the pressure.
Most classical mixed finite element methods for the Navier--Stokes equations do not preserve this property \cite{girault2012finite}.
As a consequence, the corresponding error estimates for the velocity depend on the pressure \cite{john2016finite}.
Moreover, the discrete solution $\vec u_h$ fulfills the divergence constraint
only in a discrete finite element sense.
%
This raises problems when coupling the flow simulation to a transport simulation using finite volume methods, because the maximum principle for the species concentration is directly linked to the divergence constraint in the finite volume sense \eqref{eq:discdiv0} \cite{fuhrmann2011numerical}
that is different to the finite element sense.

 The pressure robust mixed methods first introduced in \cite{linke2014role}
and comprehensively described in \cite{john2017divergence}, are based on the introduction of a divergence free velocity \textit{reconstruction operator} $\Pi$ into the discrete weak formulation of the flow problem.
The resulting discretization of the stationary Stokes equation (provided here for simplicity) reads as: find $(\vec{u}_h, p_h)\in V_h\times Q_h$ such that
\begin{align*}
  \int_{\Omega} \nu \nabla \vec{u}_h : \nabla \vec{v}_h dx   +\int_{\Omega} p_h \nabla \cdot \vec{v}_h dx
  &  =  \int_{\Omega} \vec{f} \cdot ({\Pi}\vec{v}_h) dx && \text{for all } \vec{v}_h \in V_h,\\
  \int_\Omega q_h \nabla \cdot \vec{u}_h \, dx &= 0     && \text{for all } q_h \in Q_h.
\end{align*}

This formulation differs from that of the classical mixed methods only in the right hand side. The reconstruction operator $\Pi$ has the following properties:
\begin{enumerate}
\item[(i)] If \(\vec{v}_h \in V_h\) is  divergence free in the weak sense, 
then its reconstruction $\Pi \vec{v}_h$ is pointwise divergence free: 
\begin{align*}
	\int_\Omega q_h \nabla \cdot \vec v_h \, dx = 0 \quad \text{for all } q_h \in Q_h
	\qquad \Longrightarrow \quad
	\nabla\cdot ({\Pi} \vec{v}_h) = 0 \quad \text{in } \Omega.
\end{align*}
\item[(ii)] The change of the test function by the reconstruction operator causes a consistency error that should have the same asymptotic convergence rate of the original method and should not depend on the pressure.
\end{enumerate}
Under these conditions, the resulting velocity error estimate is independent of the pressure \cite{john2017divergence}
and has the optimal expected convergence rate.
Furthermore, 
a good velocity approximation can be obtained without the need to resort to high order pressure approximations.
This allows significant reduction of degrees of freedom that are necessary to obtain a prescribed accuracy of the velocity.
The action of $\Pi$ on a discrete velocity field can be calculated locally, on elements or element patches. Therefore its implementation leads to low overhead in calculations.


\subsection{Coupling strategy.}

The coupling approach between the Navier--Stokes solver and the
Nernst--Planck--Poisson solver is based on a fixed point iteration strategy:

\begin{algorithm}[H]\SetAlgoLined \label{algo:coupling}
  Set $(\vec{u}_h, p_h)$ to zero, calculate initial solution for \eqref{eq:poisson}--\eqref{eq:incompress}\;
  \While{not converged}
  {
    Provide $\psi_h,q_h$ to Navier--Stokes solver\;
    Solve \eqref{eq:moment}--\eqref{eq:div0} for $(\vec{u}_h, p_h$)\;
    Project $\Pi\vec{u}_h, p_h$ to the Poisson--Nernst--Planck solver\;
    Solve  \eqref{eq:apoisson}--\eqref{eq:actflux}\;
  }
\end{algorithm}
The projection of the pressure to the Poisson--Nernst--Planck solver just requires the evaluation of the pressure $p_h$ in the nodes of the triangulation.

According to \eqref{eq:fluxint}, the projection of the velocity requires the integration of the reconstructed finite element solution $\Pi\vec u_h$ over interfaces between neighboring control volumes of the finite volume method.
In the implementation, these integrals are calculated by quadrature rules.
Sufficient accuracy of this step guarantees that the projected velocity is divergence free in the sense \eqref{eq:discdiv0}.
For a detailed explanation of this algorithmically challenging step, see \cite{fuhrmann2011numerical}.
The projection operator can be assembled into a sparse matrix, allowing for computationally efficient repeated application of the projection operator during the fixed point iteration.
As a consequence, in the case of electroneutral, inert transported species, the maximum principle for species concentrations is guaranteed, see \cite{fuhrmann2011numerical} for more details.
In combination with pressure robust finite element methods, this coupling approach was first applied to modeling of thin layer flow cells \cite{merdon2016inverse}.


\section{DC induced charge electroosmosis (ICEO) over an electrode with floating voltage}
\label{sect:float}

The discretization methods and the coupling strategy introduced above are implemented in the framework of the toolbox \texttt{pdelib} \cite{pdelib} that is developed at WIAS.

The solution of the Nernst--Planck--Poisson system is performed using Newton's method.
As the convergence of this method is guaranteed only in the vicinity of the solution, due to the strong nonlinearities, several measures need to be taken in order to obtain a solution.
Time embedding is an approach to solve a stationary problem from a given initial value by solving a time dependent problem with increasing time steps by an implicit Euler method.
Time step size control allows to guarantee that the difference between the solutions for two subsequent time steps is small enough to ensure the convergence of Newton's method.
Parameter ramping controls certain parameters (solvation number, surface charge) in the initial phase of the time evolution in such a way that the nonlinearities are easy to solve.
For the flow part of the problem, the stationary  Stokes problem is solved using a  second order finite element method. Its velocity space consists of piecewise quadratic
continuous vector fields enhanced with cell bubble functions and its pressure space consists of piecewise linear and discontinuous scalar fields \cite{BernardiRaugel}.
For this method, an efficient divergence free reconstruction operator into the
Raviart--Thomas finite element space of first order is constructed by standard interpolation \cite{john2017divergence,linke2016pressure}.
%
%
The method has been verified against classical Helmholtz--Smoluchowski theory and corresponding asymptotic expressions \cite{overbeek1952electrokinetic,burgreen1964electrokinetic,FGLMMPreprint}.

In order to demonstrate the capabilities of the model and the method, we simulate direct current (DC) induced charge electroosmosis (ICEO) over an electrode with a floating voltage.
The effect of induced charge electroosmosis is based on a space charge region which appears at conducting surfaces immersed in an electric field.
High electric conductivity keeps the potential of the conducting surface at a constant value.
The external application of an electric field then must result in a potential gradient at the surface which triggers the formation of a space charge region.
In this space charge region, electric forces act on the fluid, eventually setting the fluid in motion.
For a more thorough discussion of this phenomenon including the alternating current case and other numerical approaches mainly based on the Helmholtz-Smoluchowski approximation for thin double layers see e.g. \cite{squires2004induced,soni2007nonlinear,gregersen2009numerical,pimenta2018numerical}.

\begin{figure}
  \centering

  \begin{minipage}[b]{0.49\textwidth}
    \begin{tikzpicture}[scale=0.6]
      \draw[very thick]  (-4,0)--(4,0);
      \draw[very thick]  (-4,4)--(4,4);
      \draw[dashed] (-4,0)--(-4,4);
      \draw[dashed]  (4,0)-- (4,4);

      \draw[very thick, red]  (-1,0)--(1,0);

      \draw[thin,|-|] (-4,-1.0) -- (4,-1.0);
      \draw (2,-0.8) node {{\small $L$}};

      \draw[thin,|-|] (-1,-0.5) -- (1,-0.5);
      \draw (0.7,-0.3) node {{\small $L_{e}$}};

      \draw[thin,|-|] (-4.45,0) -- (-4.45,4);
      \draw (-4.75,2) node {{\small $H$}};

      \draw (-3.6,2) node {$\Gamma_l$};
      \draw (3.6,2) node {$\Gamma_r$};
      \draw (0,0.4) node {$\Gamma_{e}$};
    \end{tikzpicture}
  \end{minipage}
  \includegraphics[width=0.49\textwidth]{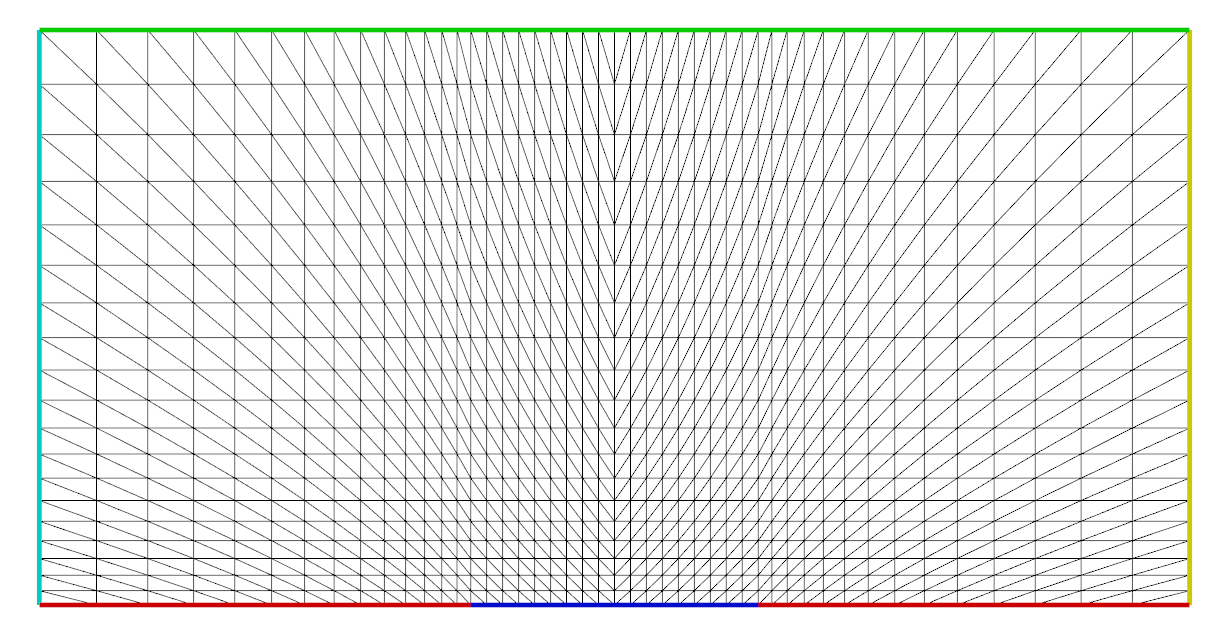}
  \caption{Left: sketch of the simulation domain $\Omega$. Right: coarse
     grid (level 1) consisting of 940 nodes}
  \label{fig:iceomodel}
\end{figure}
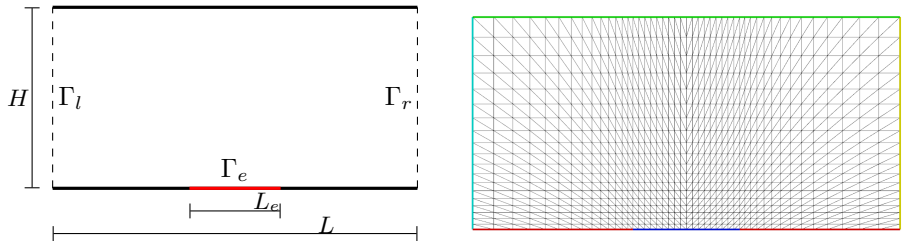

We consider two large reservoirs that are separated by an impermeable wall with a narrow channel connecting the reservoirs.
We assume that both reservoirs are filled with the same liquid electrolyte and that the ionic concentrations and the pressure are the same on both sides.
For simplicity, we assume that the channel is of infinite width with planar parallel walls.
Then, the computational domain $\Omega$ is restricted to the rectangular channel region
in a cut plane, see Figure \ref{fig:iceomodel}.
On the boundary segment $\Gamma_e$ in the middle of the bottom wall there is a metal electrode which is not connected to any outer circuit.

The channel walls on the top and on the bottom side of $\Omega$ are impermeable, i.e.\ $\vec N_i\cdot \vec n=0$ and a no slip condition is  imposed, i.e.\ $\vec u=0$.
Except at the metal electrode $\Gamma_e$, the channel walls are electrically insulating and  uncharged, i.e. $\varepsilon\nabla \psi \cdot \vec n=0$.
The metal electrode $\Gamma_e$ is assumed to consist of an ideal metallic conductor with zero resistivity
embedded in an insulating environment such that the electric potential within the metal is constant.
The value of this constant is defined by the applied electric field.
In this sense, the constant voltage at this electrode is \textit{floating}.
%
%
At the interfaces $\Gamma_l$ and $\Gamma_r$ to the reservoirs,  a zero stress boundary condition  $\nu \nabla \vec u \cdot \vec n = p \vec n$ for the flow is imposed which allows unimpeded motion of the fluid through the boundary of the simulation domain.
A partial justification of this boundary condition consists in the fact that at these boundary we can ignore the electric forces due to local electroneutrality.
Also, the same problem was simulated with a no-slip boundary condition $(\vec u=0)$ with similar results.
%
%
Since the reservoir volume is considered large compared to the channel,
we impose the rather crude approximating condition of constant prescribed
species concentrations $c_i$ at the interfaces.
The electric field is applied by imposing a potential difference between the electrolyte reservoirs.
As an approximation, bias values of the same magnitude but opposite sign
are applied at the reservoir boundaries $\Gamma_l$ and $\Gamma_r$.
Due to symmetry, it then can be assumed that the floating potential value
at the electrode on $\Gamma_e$ is zero: $\phi|_{\Gamma_e}=0$.

The potential difference between the reservoirs induces an electric current
that is carried by (positively charged) cations moving
from regions of higher potential to places with lower potential
and (negatively charged) anions moving nearly in the opposite way.
Without any charge accumulation, due to electroneutrality, there would be no influence on the velocity, and the flow velocity would be zero.

%



\begin{figure}
  \centering
  \includegraphics[width=0.49\textwidth]{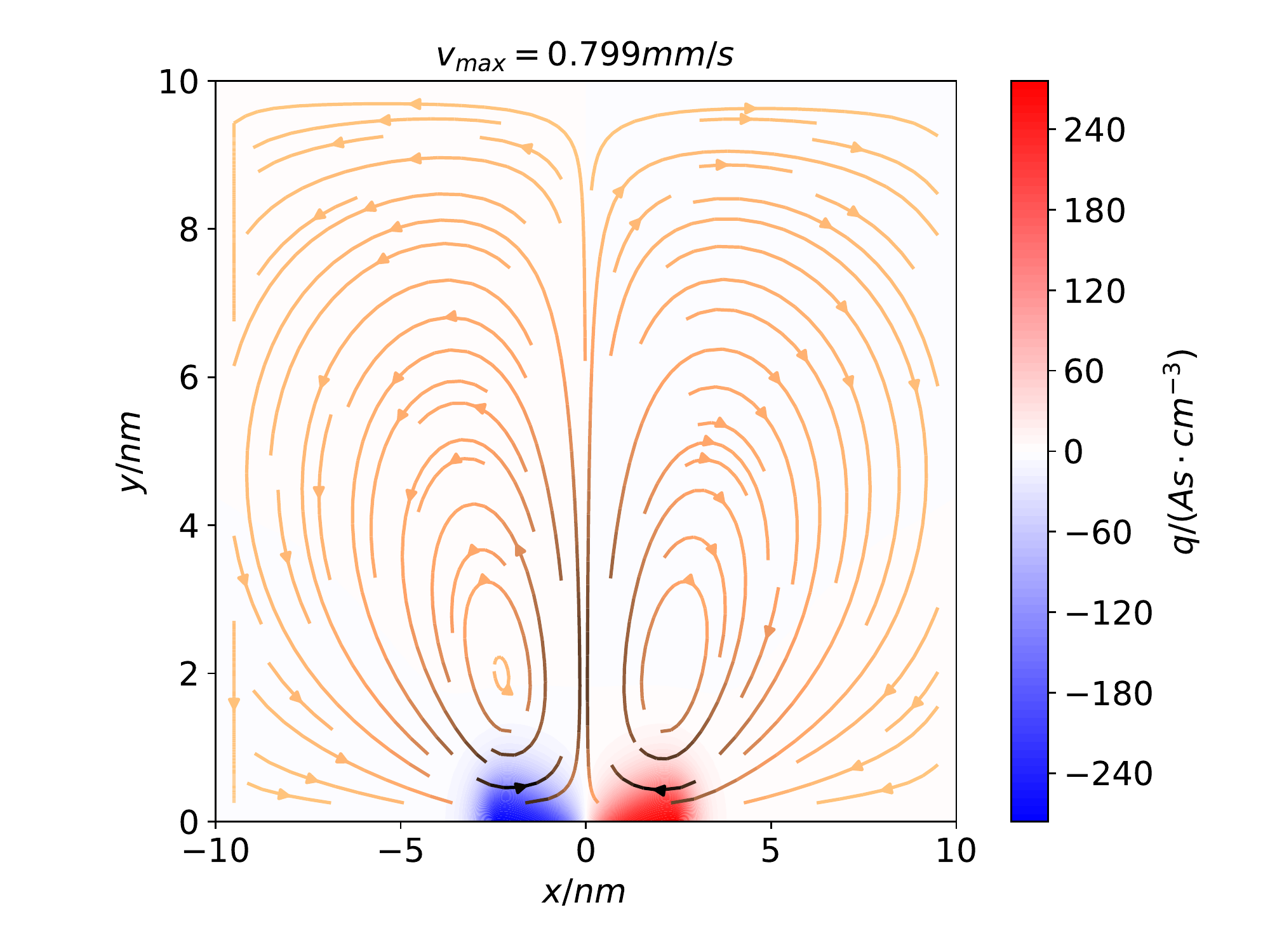}
  \includegraphics[width=0.49\textwidth]{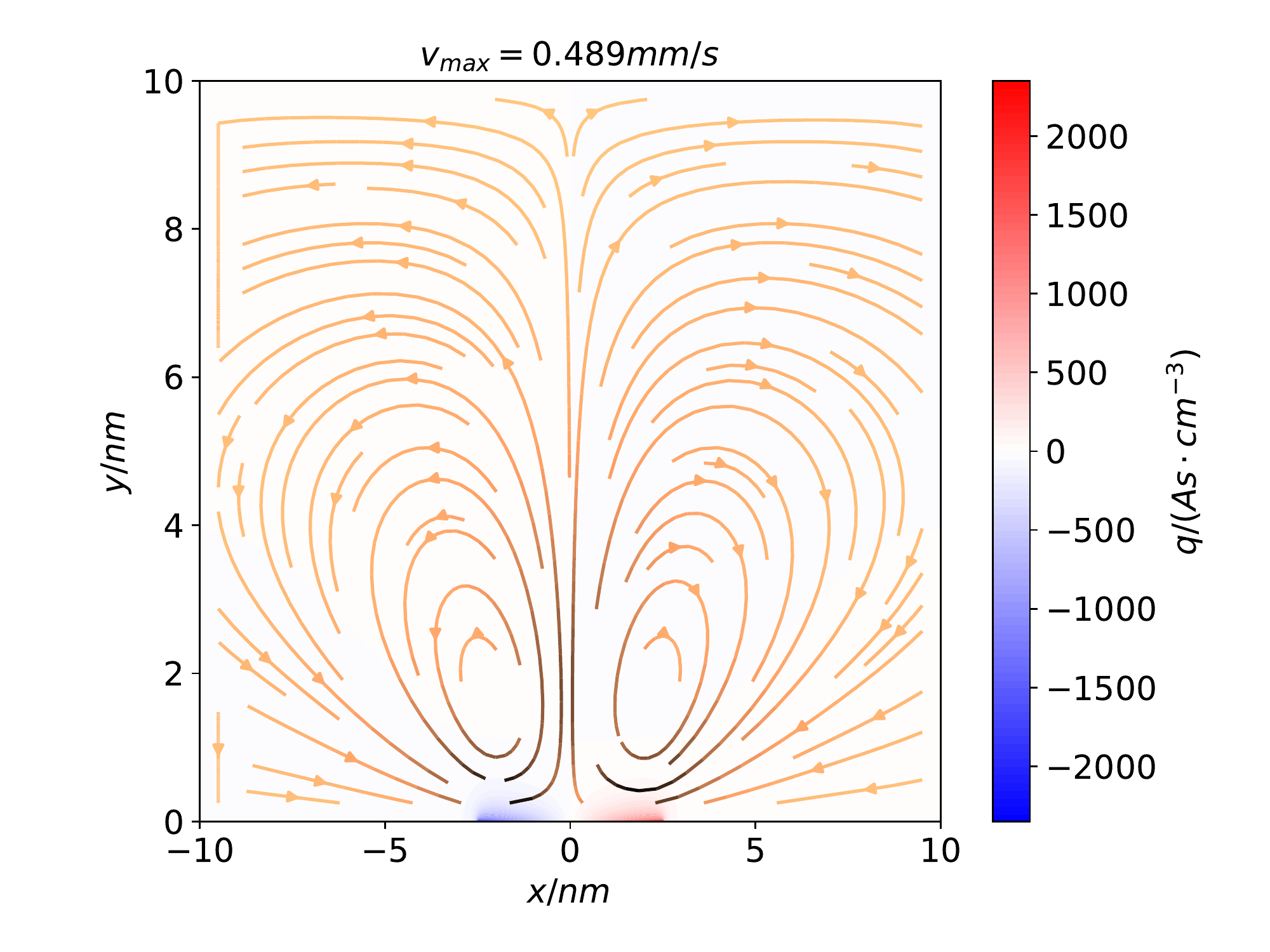}
  \caption{Charge density (color coded) and
    streamlines of the induced charge electro-osmotic flow. Left:
    Solvation model  \eqref{eq:npfull} with $\kappa=15$. Right: Classical Nernst--Planck flux \eqref{eq:classflux}. Please note the significantly different color ranges which exhibit
    the overestimation of the boundary layer charge in the case of  the classical Nernst--Planck flux.
}
  \label{fig:prof}
\end{figure}

The electric field in lateral direction triggers the formation of a polarization boundary layer at the electrode.
Due to symmetry, the charge density in this layer changes sign at the center of the electrode.
As a consequence, there are electroosmotic forces of opposite sign acting on the fluid
that lead to the creation of nanofluidic vortices.
The stream  plots in Figure~\ref{fig:prof} give a qualitative impression of the emerging flow
and charge distribution for a 1M electrolytic solution.
For the simulation, a series of discretization grids of increasing refinement level consisting of   465, 940,1875, 3690, 7596, 16383 nodes, respectively are used.
The results of the modified and classical Nernst--Planck models can be compared.
While one observes a significant difference in the width of the polarization boundary layer (due to the finite size limitations in the solvation based model), the flow pattern appears qualitatively very similar.

\begin{figure}
  \centering
  \includegraphics[width=0.49\textwidth]{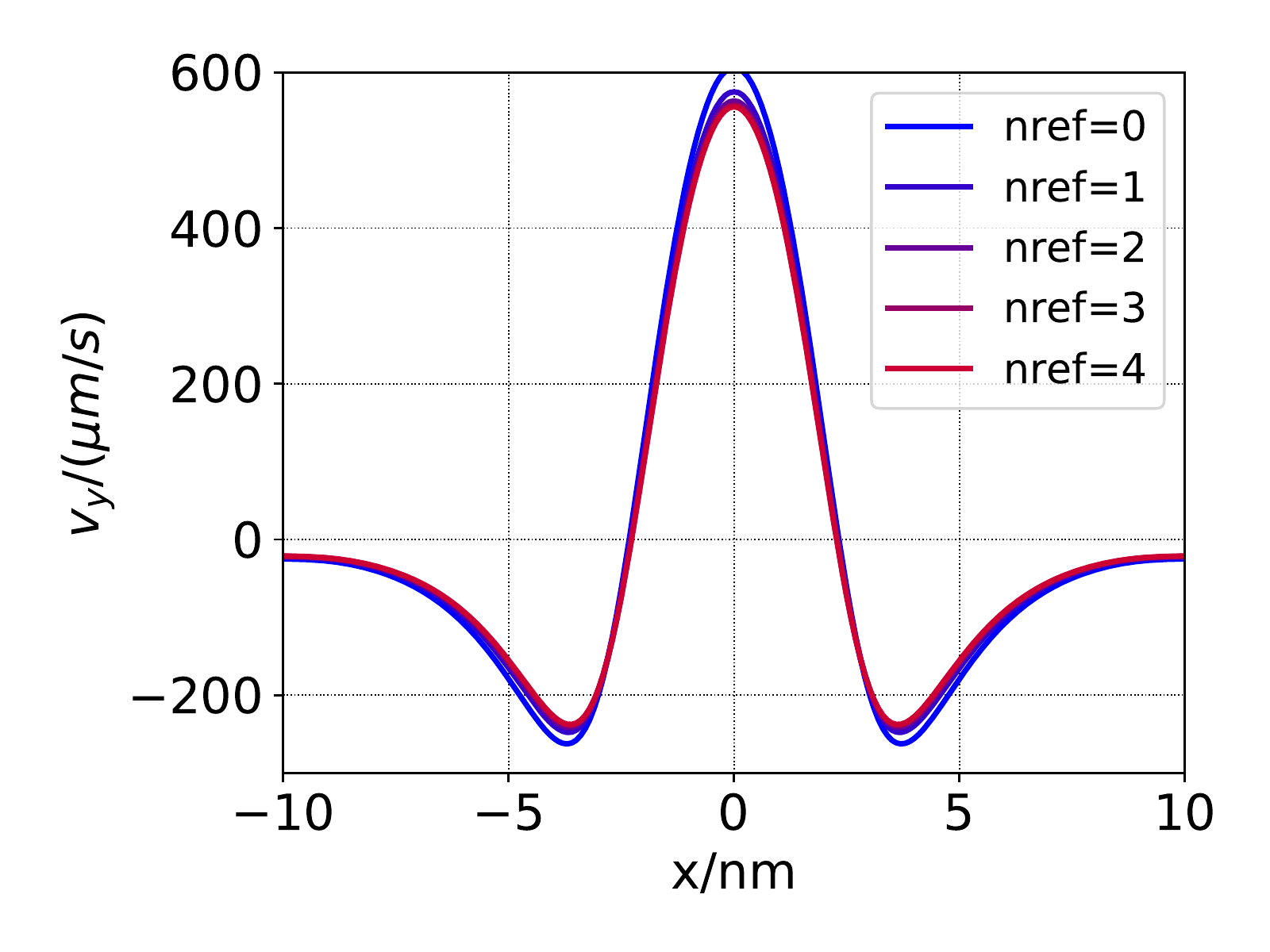}
  \includegraphics[width=0.49\textwidth]{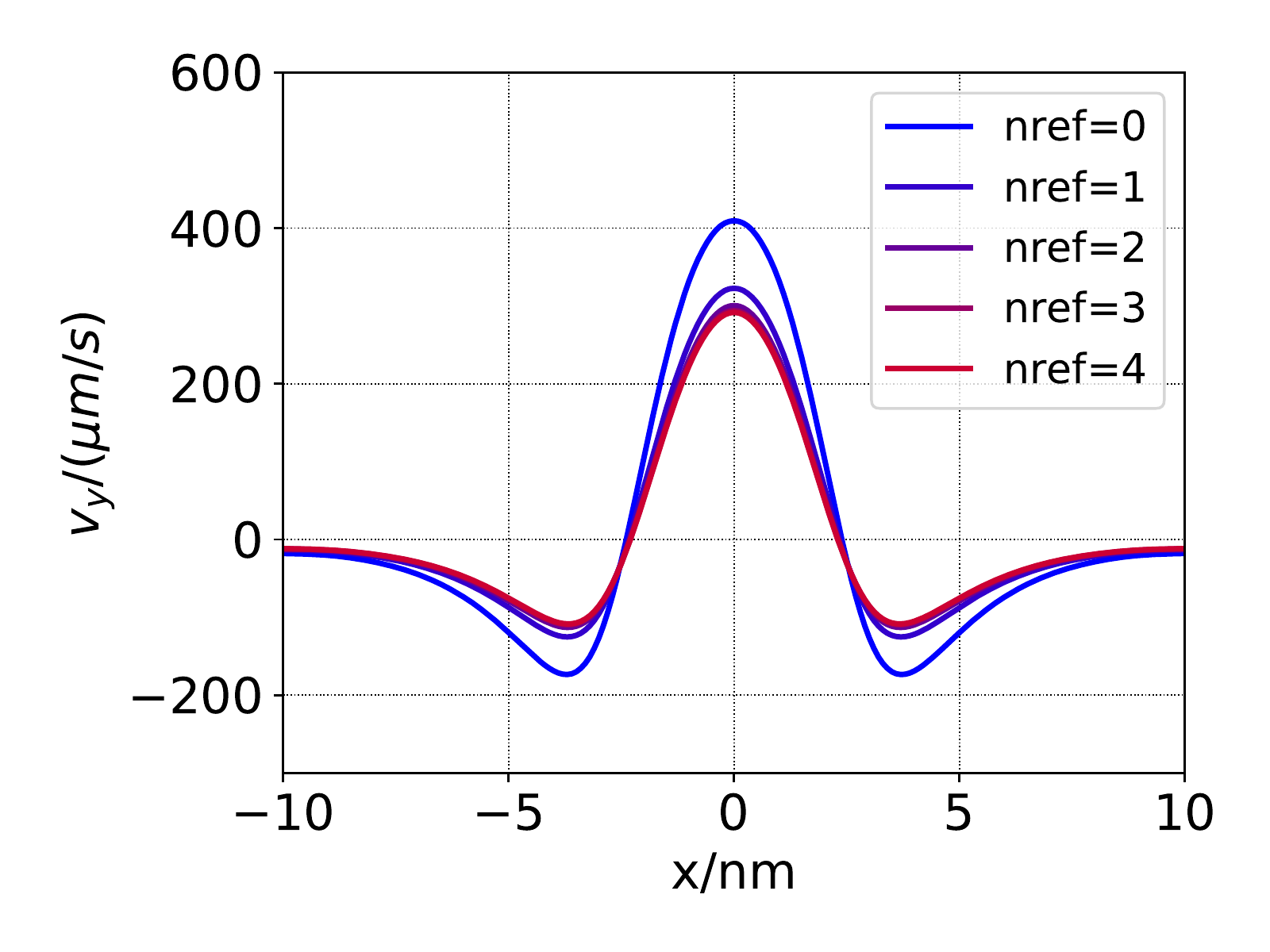}
  \caption{y-component of velocity for $y=2nm$. Left:
   Solvation model  \eqref{eq:npfull} with $\kappa=15$. Right: Classical Nernst-Planck flux \eqref{eq:classflux}.}
  \label{fig:profx}
\end{figure}

\begin{figure}
  \centering
  \includegraphics[width=0.49\textwidth]{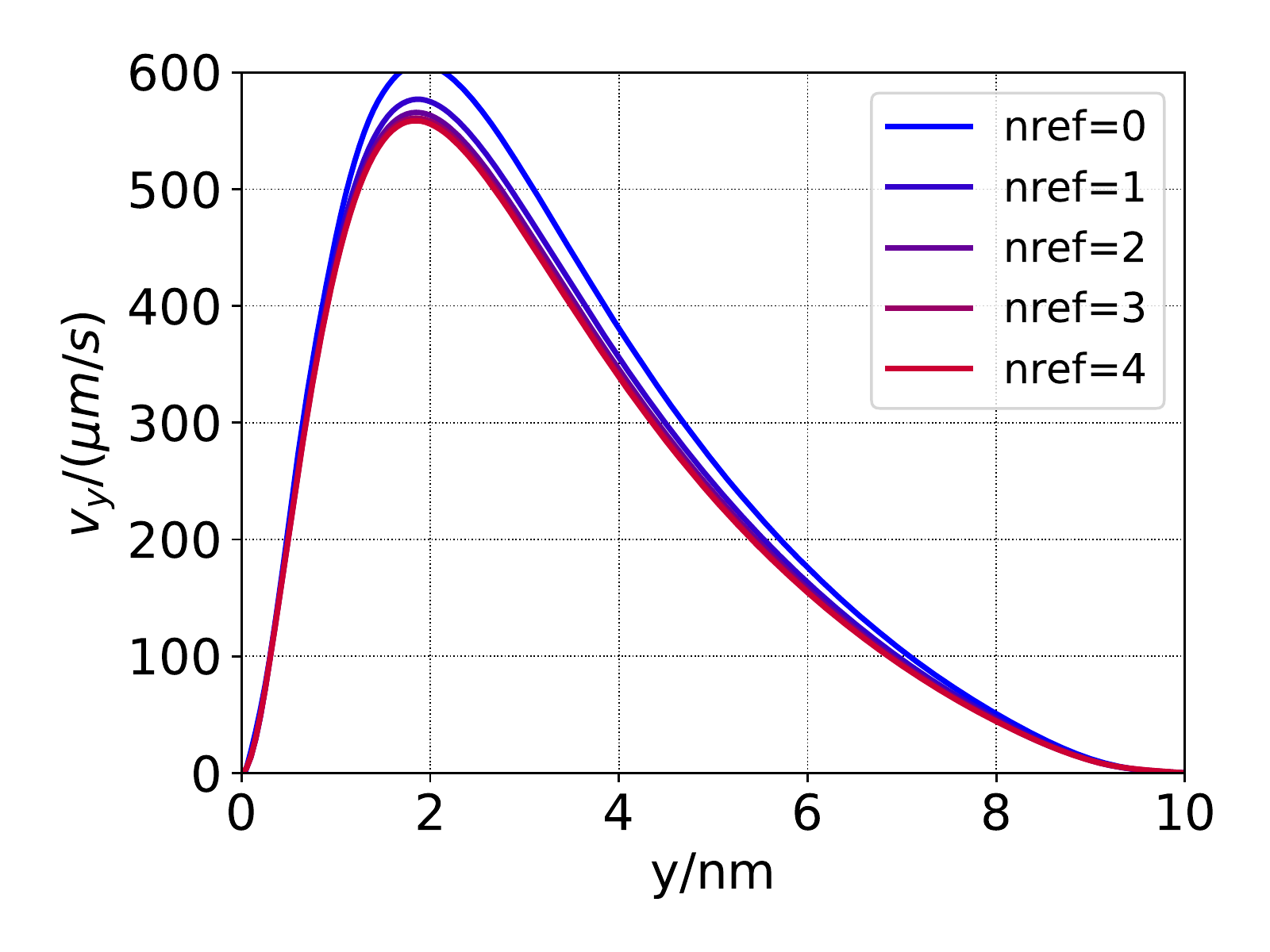}
  \includegraphics[width=0.49\textwidth]{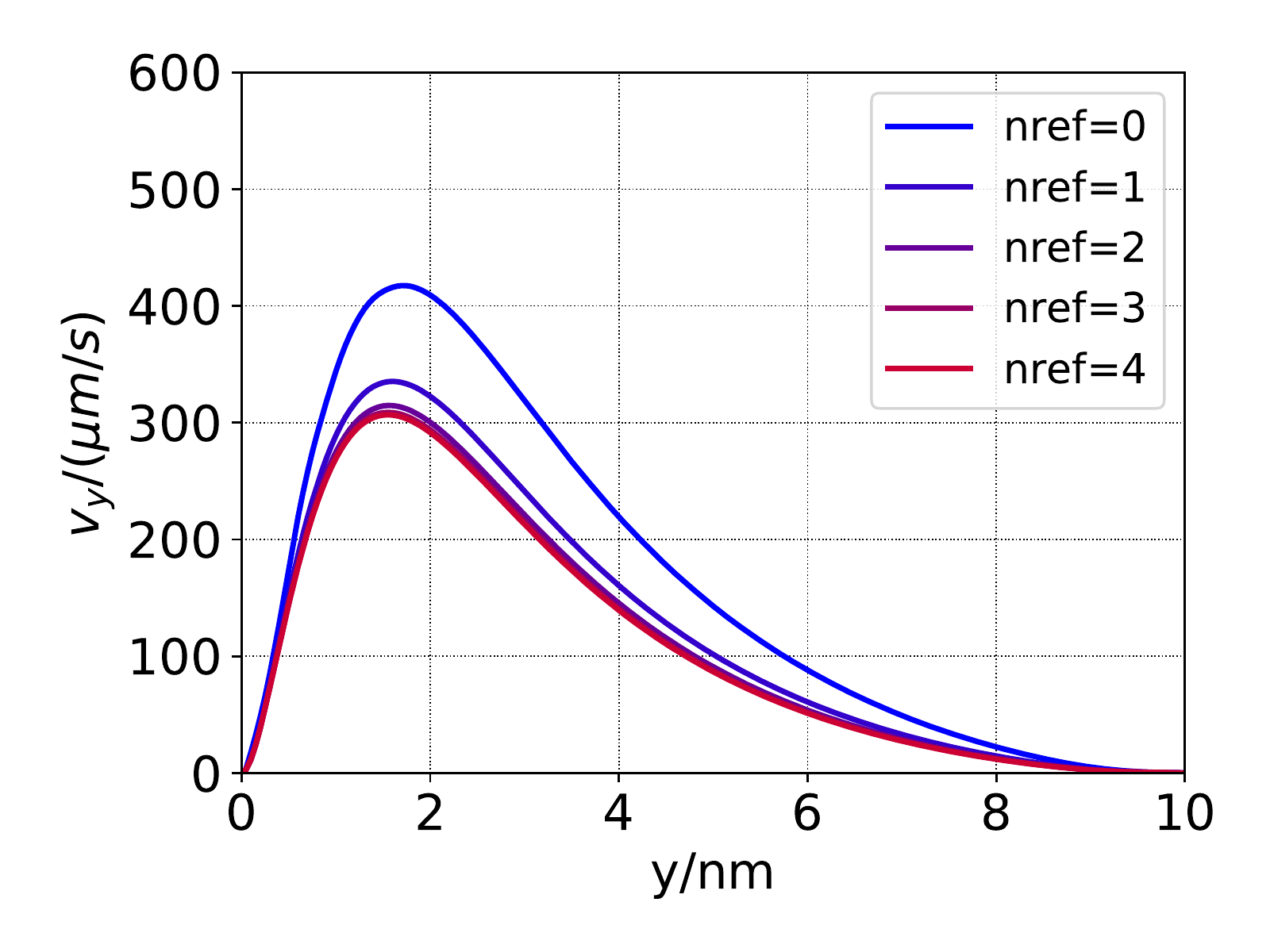}
  \caption{y-component of velocity for $x=0$. Left:
   Solvation model  \eqref{eq:npfull} with $\kappa=15$. Right: Classical Nernst--Planck flux \eqref{eq:classflux}.}
  \label{fig:profy}
\end{figure}

\begin{figure}
  \centering
  \includegraphics[width=0.49\textwidth]{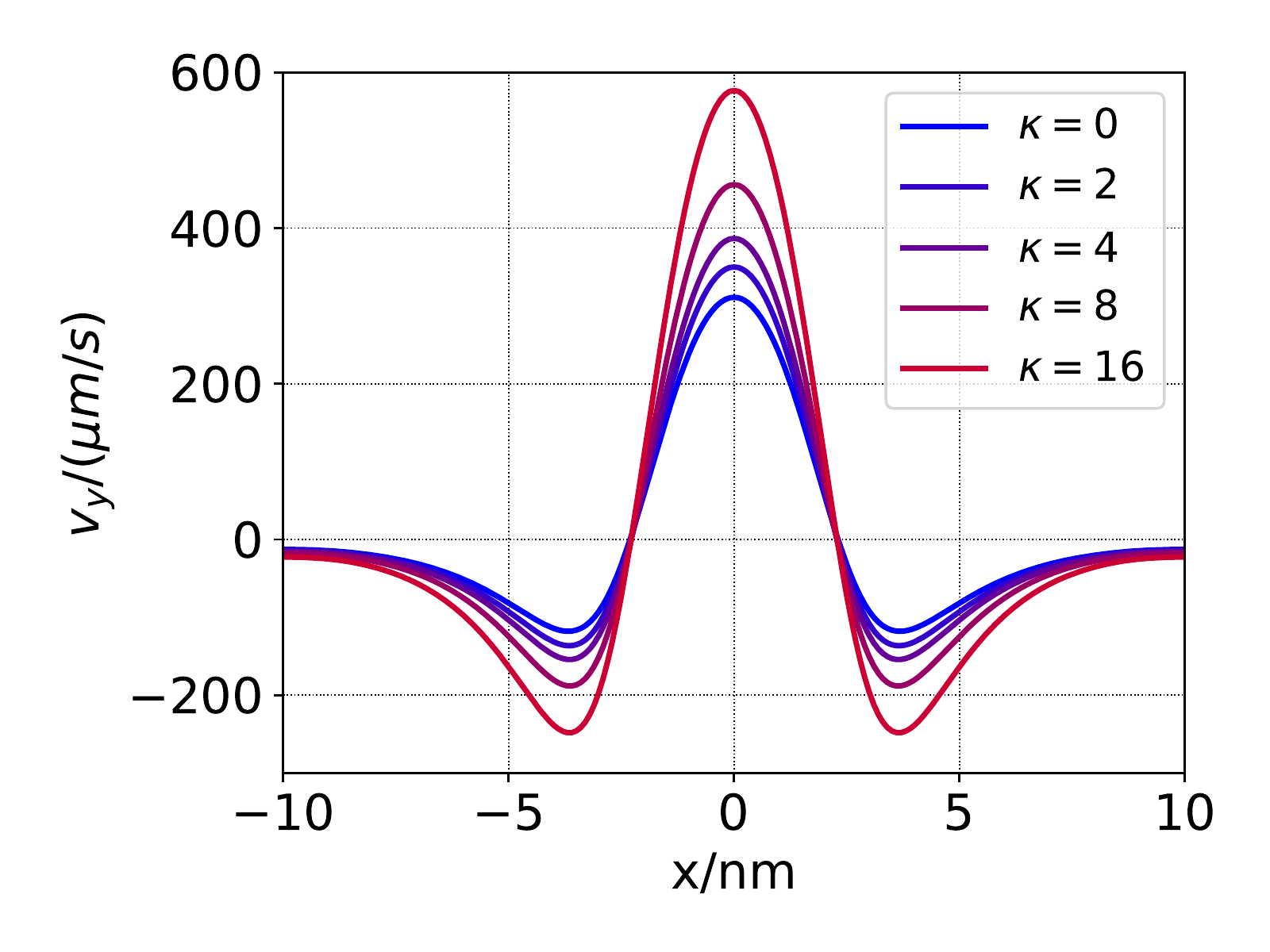}
  \includegraphics[width=0.49\textwidth]{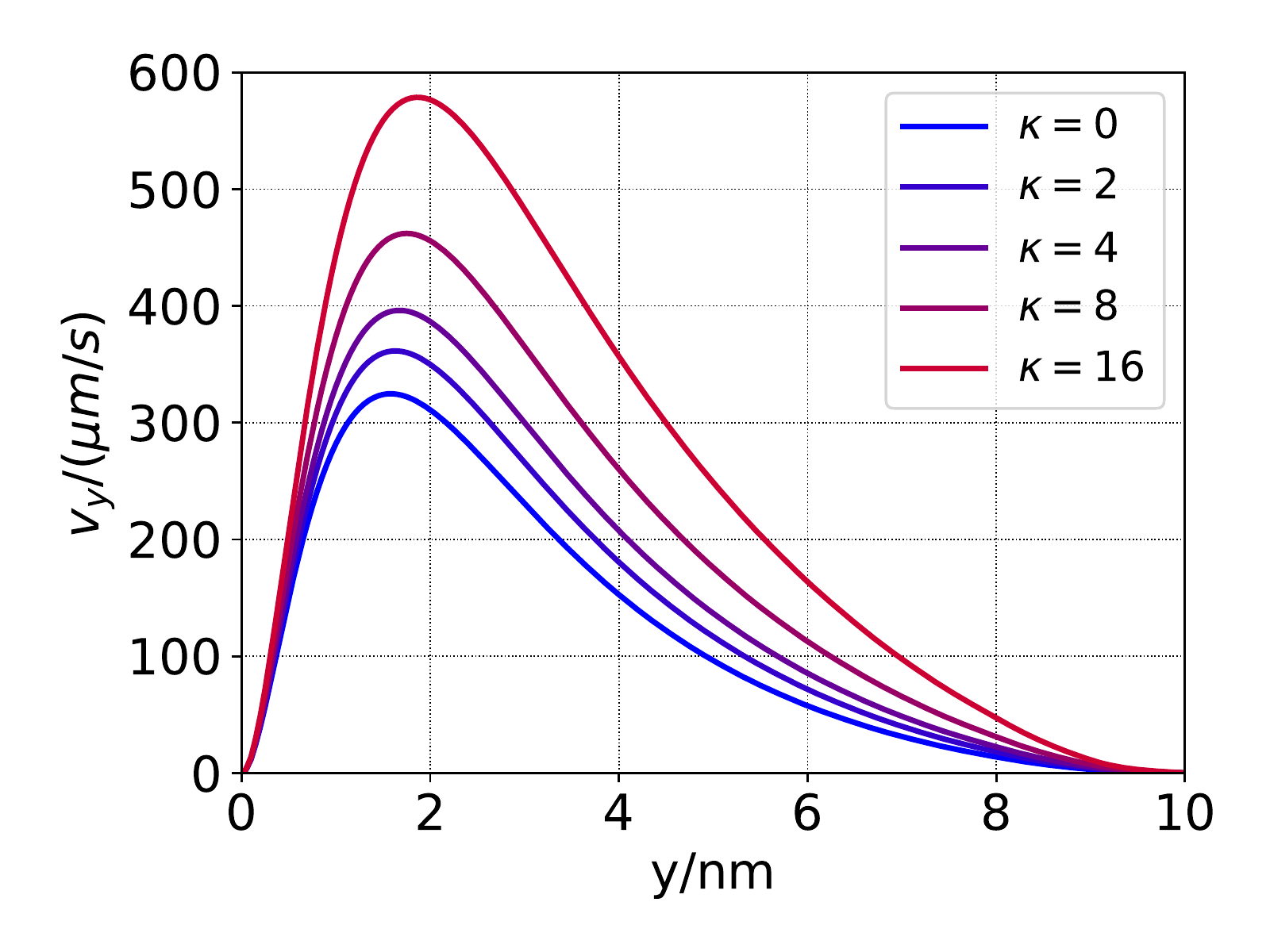}
  \caption{Velocity ($y$-component) profiles for solvation based model
    with different values of the solvation number $\kappa$.
    Left: $y=2nm$. Right: $x=0$.}
  \label{fig:profkappa}
\end{figure}

\begin{figure}
  \centering
  \includegraphics[width=0.49\textwidth]{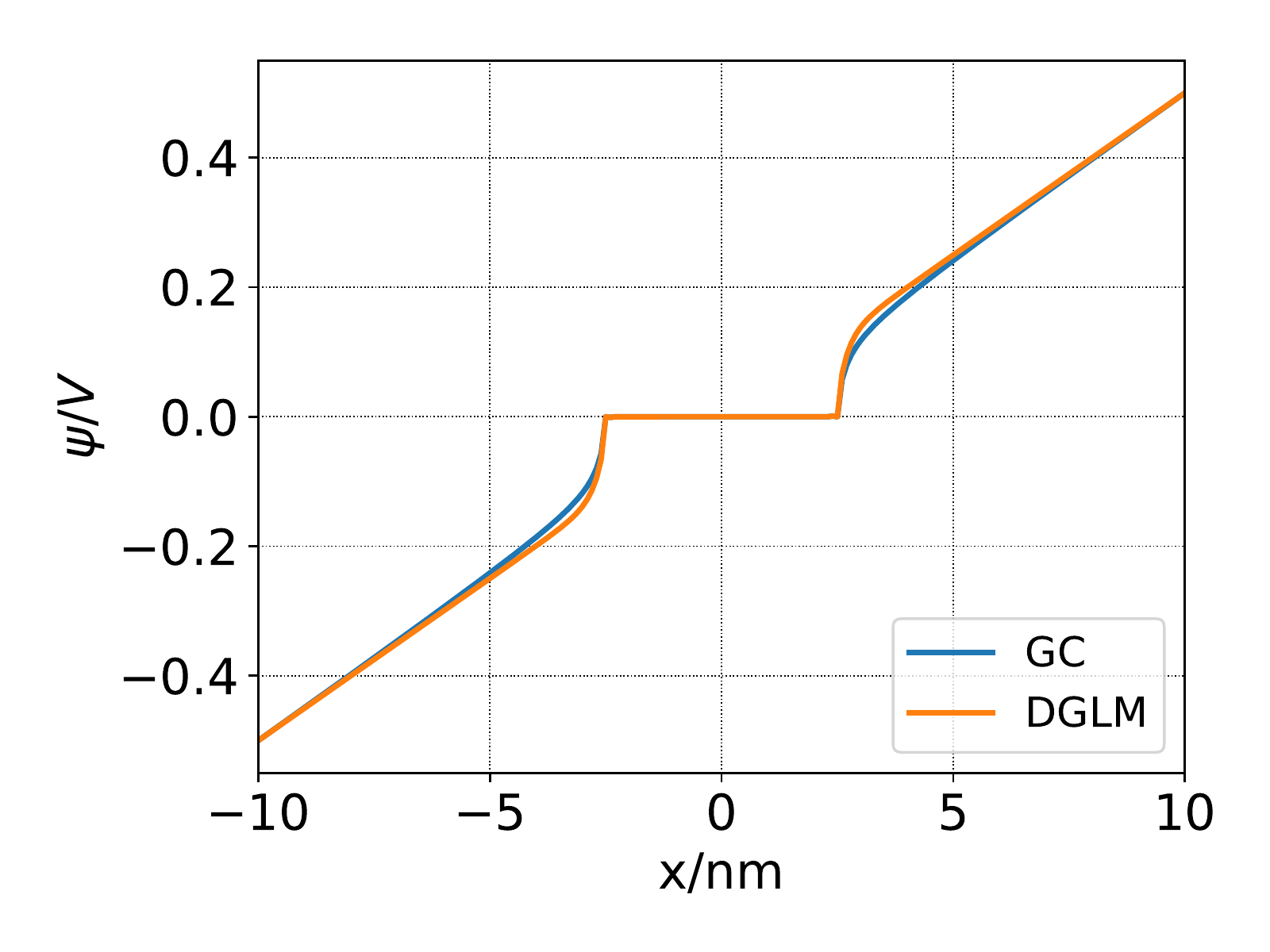}
  \includegraphics[width=0.49\textwidth]{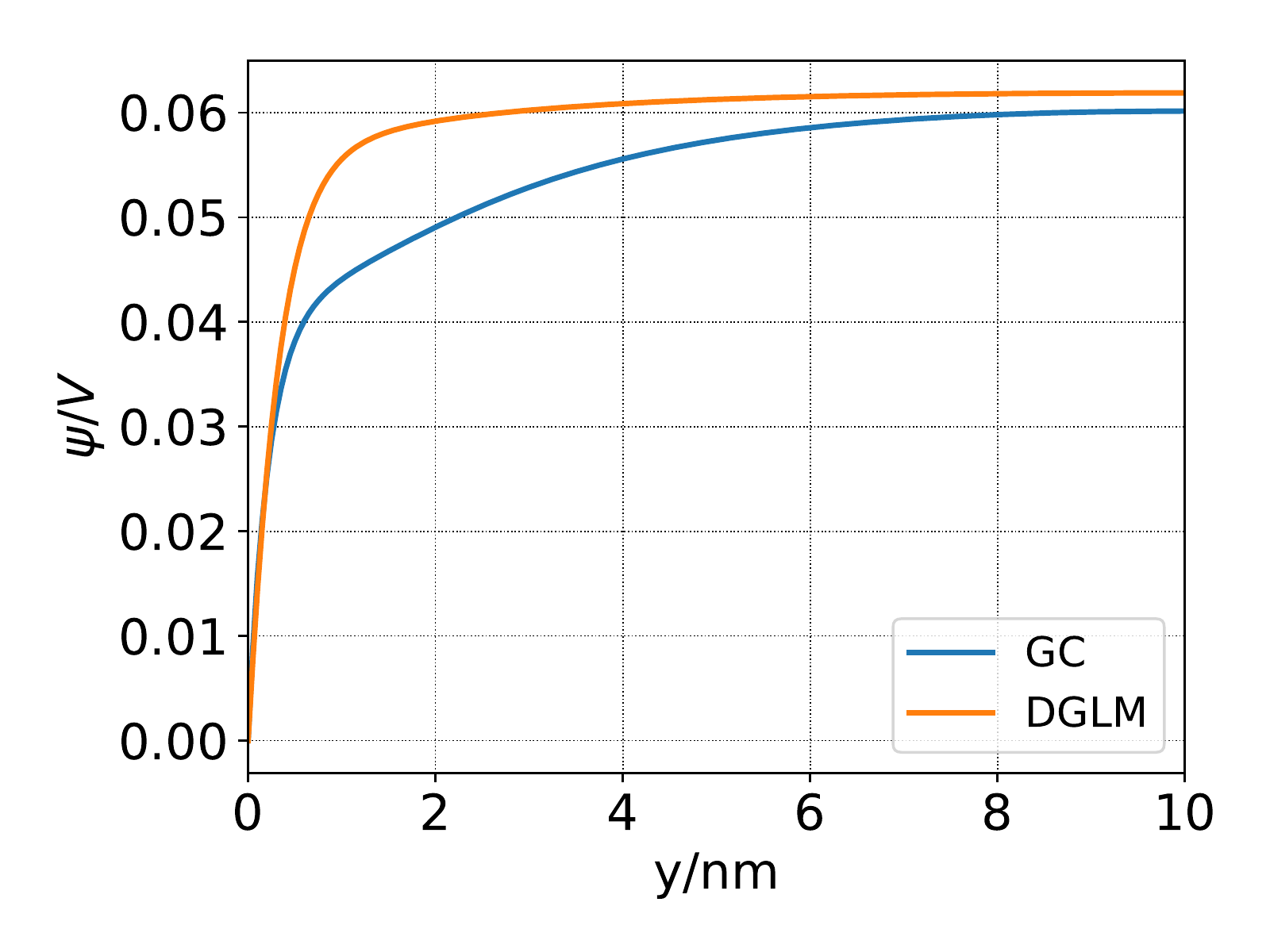}
  \caption{Profiles of electrostatic potential $\psi$ for classical Nernst-Planck (``GC'')
    and the solvation based model (``DGLM'') with $\kappa=15$.
    Left: $y=0$. Right: $x=1.25\nano\meter$.}
  \label{fig:profpot}
\end{figure}

\begin{figure}
  \centering
  \includegraphics[width=0.49\textwidth]{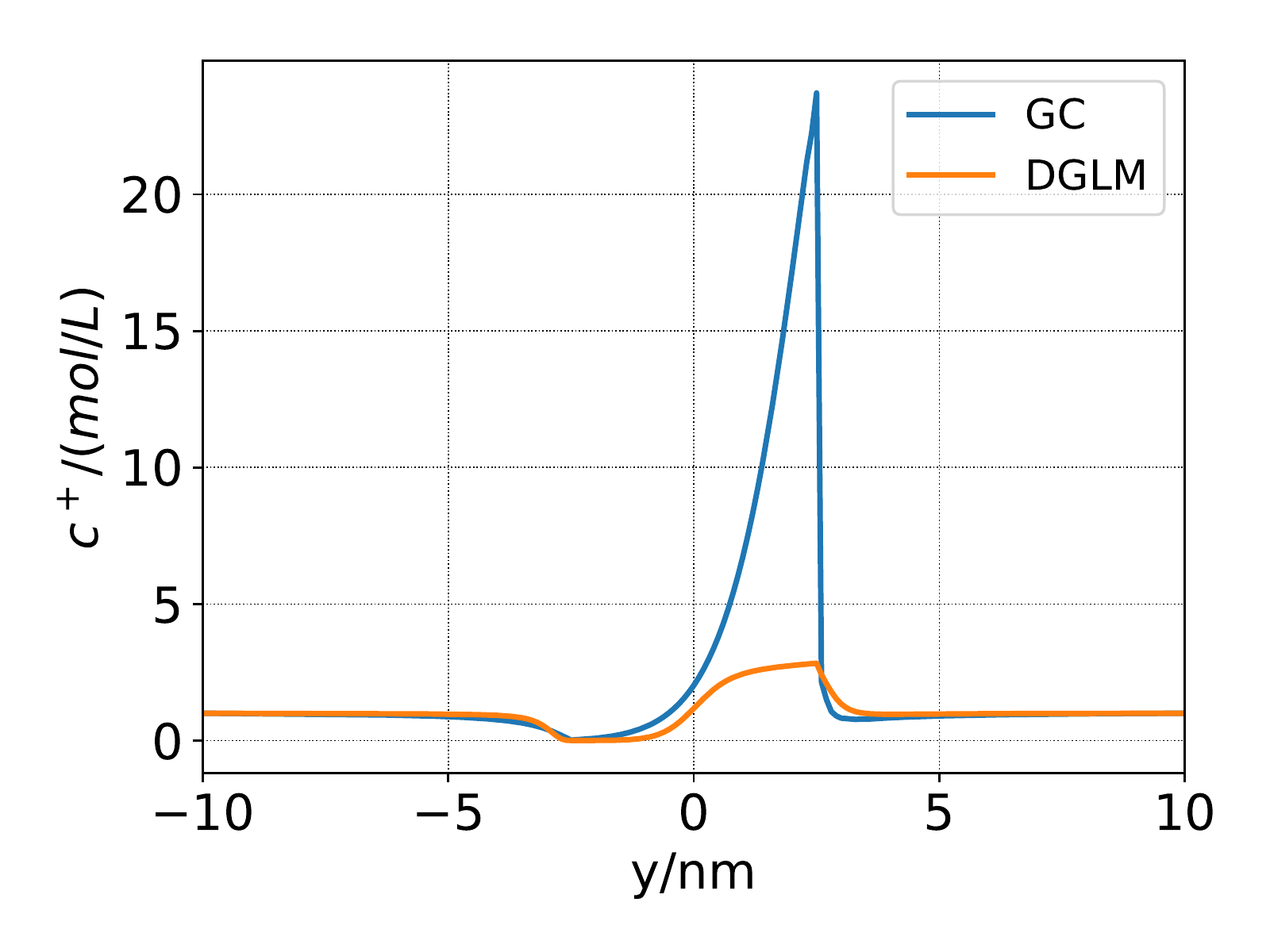}
  \includegraphics[width=0.49\textwidth]{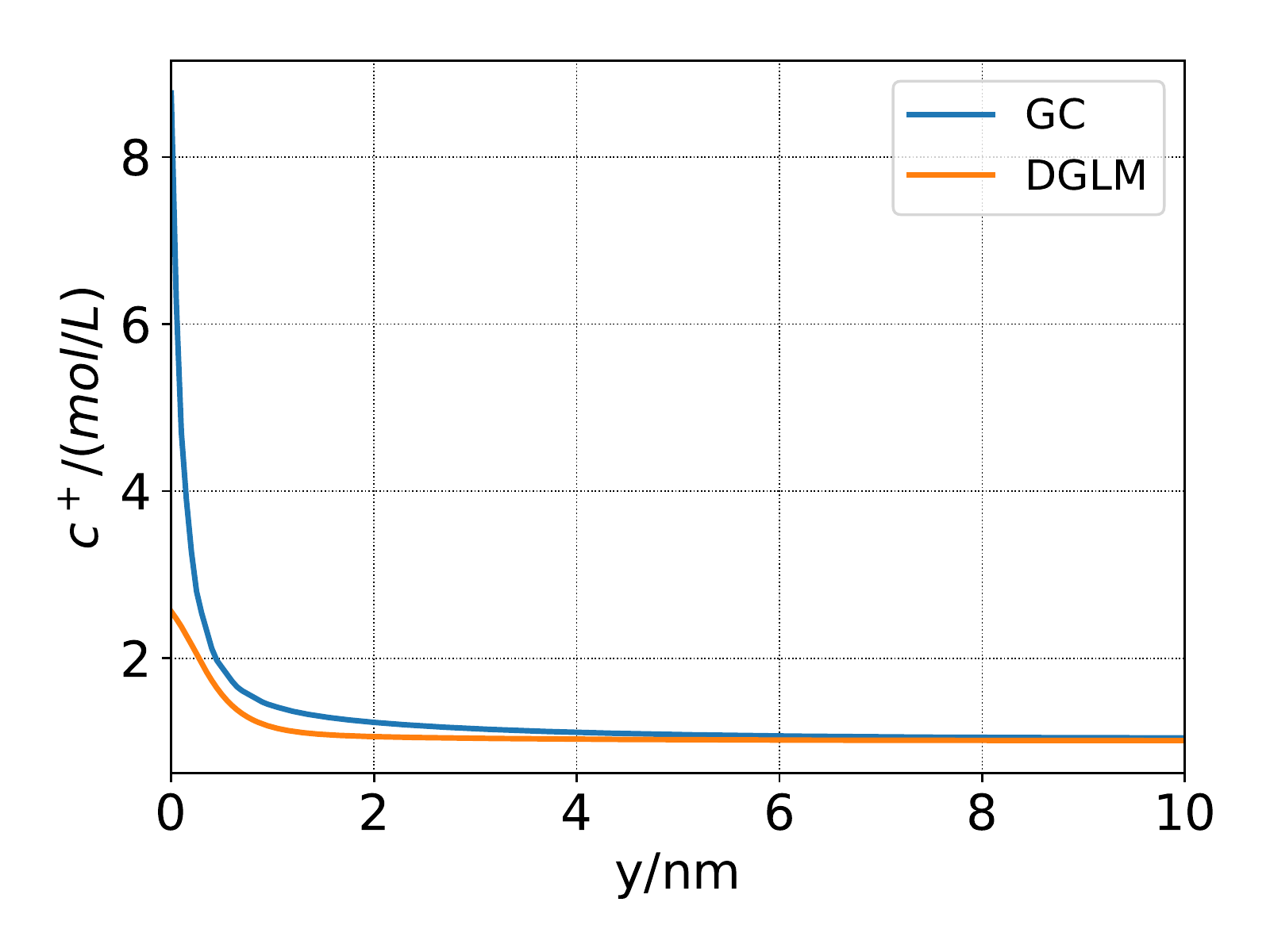}
  \caption{Profiles of positive ion concentration $c^+$ for classical Nernst-Planck (``GC'')
    and the solvation based model (``DGLM'') with $\kappa=15$.
    Left: $y=0$. Right: $x=1.25\nano\meter$.}
  \label{fig:profcp}
\end{figure}

In order to obtain a more precise picture, Figures \ref{fig:profx} and \ref{fig:profy} exhibit profiles of the $y$-component of the flow. One observes good grid convergence behavior for both kinds of models,
suggesting accurate numerical approximation of the respective solutions.
The flow velocity for the solvation based model is considerably larger than for the classical Nernst--Planck model.

Figure \ref{fig:profkappa} provides some insight into the influence of
the solvation parameter $\kappa$ on the induced flow field.
Increasing $\kappa$, a slight increase of the velocity is observed, once again consistent with the finite size effect.

Potential and concentration profiles in horizontal and in vertical direction
are given in Figures  \ref{fig:profpot} and \ref{fig:profcp}.
They demonstrate the strong differences between the classical Nernst-Planck model and the solvation based model with respect to the distribution of the electrostatic potential and the positive ions.

\section{Conclusions}
We presented a model for electro-osmotic flow at the nanoscale which includes finite ion size effects in a way that is consistent with first principles of non-equilibrium thermodynamics.
We proposed a novel numerical solution approach which preserves important structural aspects of the continuous model.
This method allows to simulate ion distributions in the vicinity of electrodes
in such a way that ion concentrations in polarization boundary layers are not drastically overestimated
and thus a meaningful coupling with the electro-osmotic flow due to the Coulomb force 
is possible.
The capabilities of the numerical method have been demonstrated by a simulation of induced charge electroosmotic flow at a nanoscale electrode with floating potential.
The simulation results for a 1M electrolyte show a considerable influence of finite ion size and solvation on electroosmotic flow velocity.

Nanoscale ion transport, electric field distribution and electrolyte flow are processes which need to be thoroughly understood in order to provide meaningful interpretations of experimental work in the emerging field of nanoscale electrochemistry.
We hope to provide valuable contributions to future research efforts in this field by amending the model with more transported species and by incorporating electrochemical reactions at electrode surfaces into the model.

\section{Acknowledgement}
The research described in this paper has been supported by the German Federal Ministry
of Education and Research Grant 03EK3027D (Network ``Perspectives for Rechargeable
Magnesium-Air batteries'') and the Einstein Foundation Berlin
within the \textsc{Matheon} Project CH11 ''Sensing with Nanopores''.

\section{Bibliography}
\bibliographystyle{elsarticle-num}

\end{document}